# Secure Zones: an Attribute-Based Encryption Advisory System for Safe Firearms

Marcos Portnoi    Chien-Chung Shen
Department of Computer and Information Sciences
University of Delaware
Newark, DE, U.S.A.
{mportnoi, cshen}@udel.edu

*Abstract*—This work presents an application of the highly expressive Attribute-Based Encryption to implement Secure Zones for firearms. Within these zones, radio-transmitted local policies based on attributes of the user and the firearm are received by embedded hardware in the firearms, which then advises the user about safe operations. The Secure Zones utilize Attribute-Based Encryption to encode the policies and user attributes, and providing privacy and security through it cryptography. We describe a holistic approach to evolving the firearm to a cyber-physical system to aid in augmenting safety. We introduce a conceptual model for a firearm equipped with sensors and a context-aware software agent. Based on the information from the sensors, the agent can access the context and inform the user of potential unsafe operations. To support Secure Zones and the cyber-physical firearm model, we propose a Key Infrastructure Scheme for key generation, distribution, and management, and a Context-Aware Software Agent Framework for Firearms.

## I. Introduction

Advances of digital technologies, as they are incorporated into devices for people's usage, often result in enhancements including ease of use, safety, precision, optimizations in resource consumption and costs. The trend has produced machines heavily dependent on cyber systems; onto these machines we rely and happily entrust our lives. Commercial and military aircraft and their complex avionics and automatic landing procedures; automobiles and their anti-lock brakes and collision avoidance systems; health patients surrending their well-being to pacemakers, electronic monitors, and robotic surgeons.

One device has persisted largely untouched by cyber technology since its inception possibly in the 12[th] century: the firearm. In light of recent tragic mass shootings, communities again claim for employing technology to boost firearm safety. In this paper, we propose evolving this mechanical device into a cyber-physical system for greater safety. In particular, we envision applying Cyphertext-Policy Attribute-Based Encryption (CP-ABE, [1], [2]) to establish zones within which, according to choice, a firearm can alert the user of unsafe operation, depending on attributes programmed in the gun's electronics.

## II. The Firearm as Cyber-Physical system

With current technology, a firearm could comprise a set of sensors enabling it to gather a wide range of information from the environment, as well as feedback outputs to the user. A software agent inside the firearm can integrate the contextual information to form an evaluation of the immediate situation; the firearm becomes context-aware. Upon this evaluation, the firearm can advise the user of potential unsafe operations through visual or haptic feedback. These sensors and feedback components are such as microphone, front- and rear-facing cameras, accelerometers, haptic and visual feedback, tamper-proof device containing the cryptography and context software agent, and radio antennas.

## III. The Secure Zone

We visualize a simple, safe schema for realizing safety areas for firearm operations within the firearm itself, with the aid of digital technologies. In this schema, a Secure Zone comprises one or more radio transmitters that will transmit a digital message containing the zone's operation policy for firearms. The message consists of an *encrypted digital signature* together with the codification of a firearm operation policy for the zone, using CP-ABE. Each Secure Zone will have its own wireless transmission with its own encrypted message representing the firearm operation policy within that zone.

Individual firearms receive, at the time of purchase or during a registration process, a private key that encodes attributes of that firearm and those of the authorized user. Within a Secure Zone range, a firearm receives the zone wireless broadcast transmission through its radio interface; embedded software agent attempts to decode the encrypted message using the firearm's private key. If successful, it means that firearm is able to safely operate within that zone; otherwise, the operation is not safe, and the firearm's feedback components will actuate to inform the user of such.

Our approach uses one-way communication, avoiding energy-costly two-way communication necessary if the firearm had to communicate with the Secure Zone radio to negotiate operations or uniquely identify a gun/user.

## IV. Key Infrastructure Scheme

This infrastructure uses CP-ABE and builds upon systems proposed by [3], [4], [5].

### A. System Model

**Central Authority (CA)**: The CA is managed by the government (the same branch that regulates firearms), and it is trusted. The CA's responsibility is to issue keys to firearms, manage the attribute set $\mathcal{A}$, and register Secure Zone Authorities, sending them cryptography parameters.

**Secure Zone Authority (SZA)**: The SZA is responsible for encrypting and wirelessly transmitting Secure Zone messages to the users/firearms. An SZA is managed by a zone administrator or owner. SZA's must register with the CA to receive proper cryptography parameters.

**Firearms (F)**: A firearm receives its ABE secret key and other security parameters upon registration with the proper government branch. This secret key expresses the attributes established by the government, according to specific rules that consider the firearm and the user/owner.

### B. Scheme Construction and Operation

**System Setup**: Central Authority generates its ABE system public/master secret key. Generates its own public/private key. Securely stores each SZA's public key and securely sends this SZA public key back to the SZA, encrypted with the CA private key, generating $CA_{PK}(SZA^i_{PubK})$, where $i$ is the particular SZA. Also sends along the system public key and a token authenticator algorithm (to prevent replay attacks) and this algorithm's system seed. The Safe Zone Authority: SZA registers with CA, receives a set of attributes $\mathcal{A}$. Generates its public/private keys. Securely sends its public key to CA, receives the system public key from the CA and generates $CA_{PK}(SZA^i_{PubK})$.

**User Registration and Secret Key Generation**: upon registration and user authentication with the government branch, a set of attributes $\mathcal{A}_u$ will be determined for the user $u$. The CA will generate the firearm/user ABE secret key expressing the user attributes (bonded by a random number $x$ to prevent colluding) together with the firearm ID, user ID, $x$ and expiration time $et$. This key is recorded into the firearm's TPD, along with the token authenticator algorithm, its system seed, and the CA's public key.

**Secure Zone Authority Encrypted Transmission**: $SZA^i$ decides on a Secure Zone policy $f$ over the attribute set $\mathcal{A}$. Generates token $tk$ (from token authenticator algorithm) and hashes it using standard hashing algorithm; obtains $hash(tk)$. Encrypts $hash(tk)$ with the SZA private key, generating $SZA^i_{PK}(hash(tk))$. Uses the token $tk$ as symmetric key to encrypt $SZA^i_{PK}(hash(tk))$ together with $CA_{PK}(SZA^i_{PubK})$, and a timestamp $ts$. Uses the ABE system public key to encrypt previous result, $tk(SZA^i_{PK}(hash(tk)), CA_{PK}(SZA^i_{PubK}), ts)$, per the policy $f$; the final result is the message $m_{SZA^i}$. This message is broadcast via radio following a certain wireless communications standard.

**Decryption, Secure Operation Assessment**: a firearm $u_i$ within the range of $SZA^i$ Secure Zone receives, via its antenna, the message $m_{SZA^i}$. The software agent applies the decryption secret key. If the firearm has the appropriate attribute set $\mathcal{A}_{u_i}$ such that $f(\mathcal{A}_{u_i}) = 1$, then the decryption is successful, and the agent obtains $tk(SZA^i_{PK}(hash(tk)), CA_{PK}(SZA^i_{PubK}), ts)$. Agent runs its own token authenticator algorithm (with the same seed as the SZA, recorded at registration), obtains $tk_u$. Agent uses $tk_u$ as symmetric key to decrypt last outcome and get $ts$, $SZA^i_{PK}(hash(tk))$, and $CA_{PK}(SZA^i_{PubK})$. If $tk_u \neq tk$, decrypt will fail; stop and alert user. Else, if $et < ts$ (secret key expired), then stop and alert user. Else, use the CA public key stored with the agent to decrypt $CA_{PK}(SZA^i_{PubK})$. Use the resulting $SZA^i_{PubK}$ to decrypt $SZA^i_{PK}(hash(tk))$; if final product is different than $hash(tk_u)$ (apply the same standard hashing algorithm to $tk_u$, compare to $hash(tk)$), then at least one of the previous keys is invalid, or the message is invalid. Stop and alert user. Else, firearm is authorized to operate within this Secure Zone. If, in the first step, $f(\mathcal{A}_{u_i}) \neq 1$, then decryption is unsuccessful, resulting in a message of invalid format. Stop and alert user.

The infrastructure will also provide for key revocation and expiration, and key update.

## V. Preliminary Experimentation and Future Work

Our initial experimentation focuses on implementing the key infrastructure and performing simulations, in which we evaluate and demonstrate whether mobile users within secure zones range can properly receive and decrypt the messages, if they have the appropriate attributes. Our future work involves investigating the effectiveness of the proposed security model, and how to best address the issues of key revocation and update, as well as its performance in light of several attack vectors. Moreover, we will augment our model with delegating key generation to the Secure Zone Authorities.